\documentclass[final,5p,times,twocolumn]{elsarticle}

\usepackage{dblfloatfix}
\usepackage{euscript}   
\usepackage{graphicx}
\usepackage{dcolumn}
\usepackage{bm}
\usepackage{amsmath}

\usepackage{textcomp}
\usepackage{tabularx} 
\usepackage{times}
\usepackage{wrapfig}
\usepackage[table]{xcolor} 
\usepackage{caption}
\usepackage{subcaption}
\usepackage{adjustbox}
\usepackage{listings}
\usepackage{silence}
\usepackage{wasysym}
\usepackage{titlesec}
\usepackage{hyperref}
\usepackage{url}
\hypersetup{
    colorlinks=true,
    linkcolor=green,
    filecolor=magenta,      
    urlcolor=cyan}

\usepackage{amssymb}
\usepackage{amsthm}

\usepackage{diagbox}

\journal{Nanotechnology}

\begin{document}

\begin{frontmatter}


\title{Computer Simulation of Carbonization and Graphitization of Coal}

\author[OU]{C. Ugwumadu \corref{cor} \fnref{lb2}}
\ead{cu884120@ohio.edu}
\author[CONSOL]{R. Olson III}
\author[CONSOL]{N. L. Smith}
\author[OU]{K. Nepal}
\author[RUSS]{Y. Al-Majali}
\author[RUSS]{J. Trembly}
\author[OU]{D. A. Drabold \corref{cor}}
\ead{drabold@ohio.edu}

\address[OU]{Department of Physics and Astronomy,
Nanoscale and Quantum Phenomena Institute (NQPI)\\
Ohio University, Athens, OH 45701, USA}

\address[CONSOL]{CONSOL Innovations, Triadelphia, WV 26059, USA}

\address[RUSS]{Department of Mechanical Engineering,
Institute for Sustainable Energy and the Environment (ISEE), \\
Ohio University, Athens, Ohio 45701, USA.}

\cortext[cor]{Corresponding Authors}

\fntext[lb2]{C.U. served as a Research Intern at CONSOL Innovations LLC during the preparation of this manuscript.}

\begin{abstract}
This study describes computer simulations of carbonization and graphite formation, including the effects of hydrogen, nitrogen,
oxygen, and sulfur. We introduce a novel technique to simulate carbonization, ``Simulation of Thermal Emission of Atoms
and Molecules (STEAM),” designed to elucidate the removal of volatiles and density variations in carbonization residue. The
investigation extensively analyzes the functional groups that endure through high-temperature carbonization and examines the
graphitization processes in carbon-rich materials containing non-carbon “impurity elements”. The physical, vibrational, and
electronic attributes of impure amorphous graphite are analyzed, and the impact of nitrogen on electronic conduction is investigated,
revealing its substitutional integration into the sp$^2$ layered network.

\end{abstract}

\begin{keyword}
coal, carbon, amorphous graphite, molecular simulation, carbonization, graphitization
\end{keyword}

\end{frontmatter}



\section{\label{sec:introduction}Introduction}

The global graphite shortage has steered research towards alternatives to natural graphite for current and emerging technologies \cite{graphitedeficit}. Among these materials, coal has garnered natural attention. Although the detrimental effects of coal combustion for energy generation on health, climate, and the environment have been extensively documented in previous studies \cite{c2,c3,c4}, contemporary apprehensions regarding the scarcity of natural and synthetic graphite (currently on the list of critical materials \cite{news4}) have prompted a fresh appraisal of coal's potential as a carbon resource \cite{rebranding, news1, news3}. In the United States, research on coal utilization has extended beyond energy production to include using coal-based cryptocrystalline graphite as electrode materials in lithium-ion \cite{shi2021coal, bora2021coal, liu2010microstructure} and aluminum-ion \cite{song2019cellulose} batteries.  Additionally, there is a growing interest in using coal as filler material in composites to enhance the mechanical properties of polymers \cite{Yahya}, as well as to improving the electrical properties of metals \cite{news2}. The later exemplified by successful cases using polycrystalline graphene composites with copper \cite{K1} and aluminum \cite{k2}.

Synthetic graphitization of coal begins with carbonization - the controlled heating of coal within the temperature range of 800 K to 1500 K, in the absence of oxygen. This elevated temperature triggers the release of volatile compounds including gaseous hydrocarbons, oxides of carbon, organo-nitrides, and organo-sulfides, yielding a solid residue known as coke \cite{coalToCoke}. The consensus is that subjecting the coke to higher temperatures (2500 K $\sim$ 3200 K) induces the cleavage of aliphatic chains and the formation of polyaromatic compounds through free radical crosslinking \cite{crosslinking1}. These polyaromatic compounds exhibit interconnected carbon atom rings, resulting in structures having sp$^2$ hybridization, akin to graphite \cite{coalPyrolysis}. However, research suggests that non-carbon elements in graphitic precursors (e.g., coal) such as hydrogen, nitrogen, oxygen, and sulfur, remain present in the final graphite material and interfere with its formation \cite{landmark1,carbo_graphi,irreversible}. For instance, Franklin demonstrated that the presence of oxygen can render the graphite precursor non-graphitizing \cite{franklin}. Bourrat and co-workers showed that sulfur can endure in the coke after carbonization of some graphite precursors for temperatures above 2000 K, forming crosslinks between the graphitic layers \cite{bourrat}. 

Established experimental techniques such as Raman Spectroscopy \cite{Raman}, Fourier transform infrared spectroscopy \cite{FTIR}, electron paramagnetic resonance \cite{ESR}, and others, can inform on the structure of the materials obtained at different stages from coal carbonization to its graphitization, but they fall short in clarifying how low concentrations of non-carbon atoms influence properties like the electronic conduction in coal-derived nanostructures. 

Molecular simulation is an ideal method to gain insight into the atomistic attributes of coal and its derived nanostructures. Recent studies led by Thapa and colleagues \cite{aG, aG2} have provided a new perspective on the graphitization mechanism of carbonaceous materials, painting an atomistic portrait of amorphous graphite formation at elevated temperatures in elemental carbon. This work challenged the notion that layering in graphite emerged solely from Van der Waals forces, and indicated that layering is due significantly to wavefunction mixing, involving interactions among $\pi$ electrons in the galleries. This insight, also observed in other layered carbon allotropes such as multi-shell fullerenes \cite{aF} and multi-walled nanotubes \cite{aCNT}, emphasizes the importance of molecular simulations to understand the chemistry and bonding in complex systems.

Our study examines carbonization and the emergence of layered nanostructures derived from coal-like computer models. We include carbon, hydrogen, oxygen, nitrogen, and sulfur. A new method for simulating carbonization, known as the "Simulation of Thermal Emission of Atoms and Molecules (STEAM)," is presented and explored in detail. Analysis of gas evolution from  bituminous coal models during carbonization, and the functional groups present in the coke are conducted.  Additionally, \textit{ab initio} density functional theory (DFT) \cite{DFT} is employed to probe graphitization in models with ``coal-like" elemental compositions containing 5\% and 10\% non-carbon constituents (by weight). The analysis encompasses an exploration of the vibrational modes, electronic structure, and transport in impure amorphous graphite. Furthermore, the study investigates the impact of carbon-substituted nitrogen on the electronic transport in the layered amorphous nanostructure, utilizing the space-projected conductivity technique \cite{SPC,pssb_2020}.

\section{Computational Methods}

Molecular dynamics (MD) simulations were conducted using two methods: (1) The reactive force field (REAXFF) inter-atomic potential \cite{CPotReaxff}, including hydrogen, carbon, nitrogen, oxygen, and sulfur, and (2) DFT with plane-wave pseudo-potentials for all the aforementioned elements except hydrogen. The REAXFF calculations were executed within the Large-scale Atomic/Molecular Massively Parallel Simulator (\texttt{LAMMPS}) \cite{lammps}, while the Vienna \textit{Ab initio} Simulation Package (\texttt{VASP}) \cite{VASP} was employed for DFT computations. A single \textbf{k}-point ($\Gamma$) and periodic boundary conditions were used for all DFT calculations. When relevant, simulations followed the canonical or the isobaric-isothermal ensemble at specified temperature and/or pressure, controlled by a Nos\'e-Hoover thermostat and/or barostat \cite{Lammps_NPT1,Lammps_NPT2,Lammps_NPT3}. MD simulations using REAXFF and DFT employed timesteps of 0.25 fs and 1.5 fs, respectively. The subsequent sections elaborate on the specific simulation protocols for carbonization and graphitization.

\section{Results \& Discussion}

\subsection{Carbonization \label{sec:reaxff}}

\begin{figure}[h!]
	\centering
	\includegraphics[width=\linewidth]{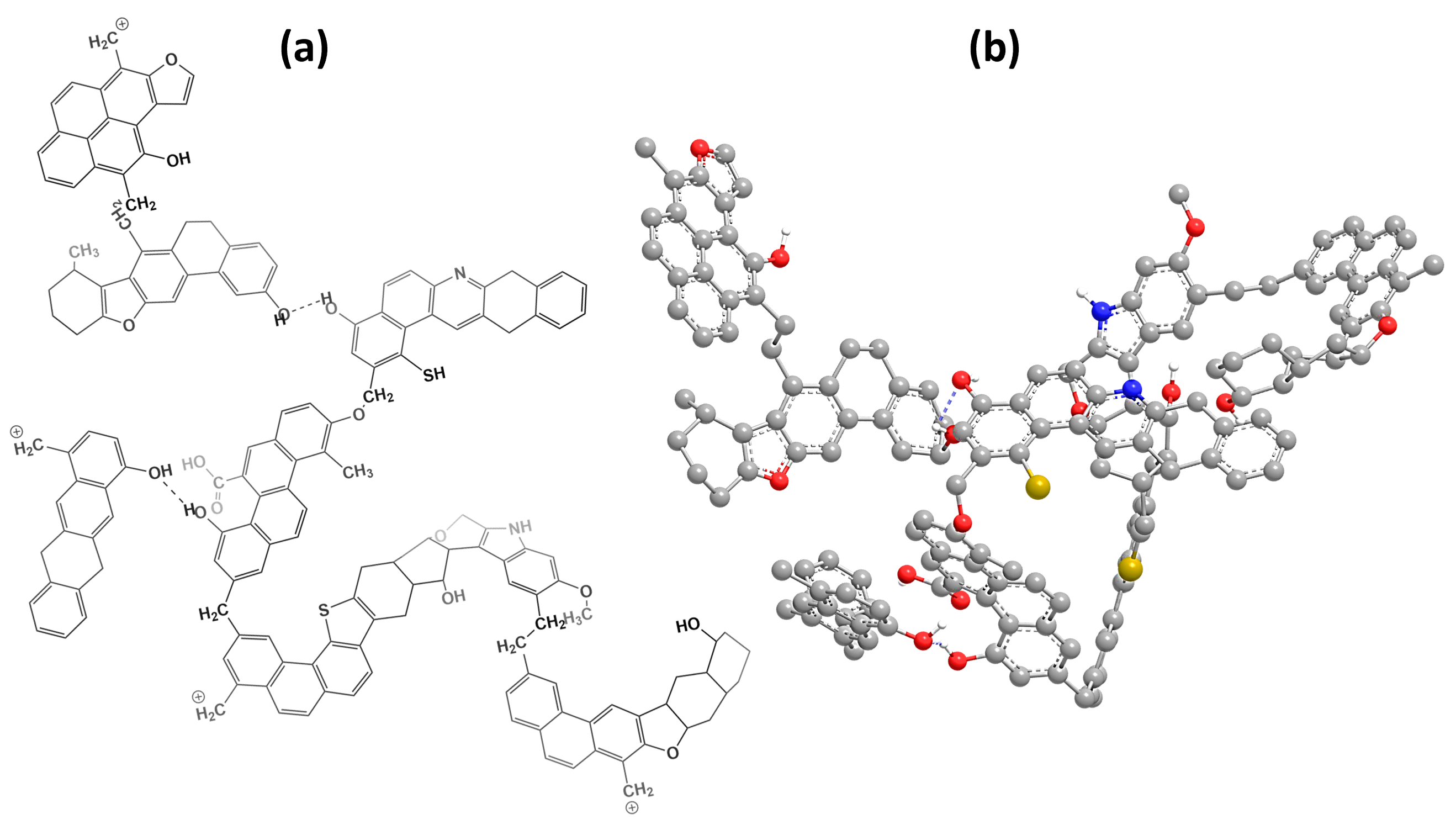}
	\caption{(a) The 2D coal model proposed by Solomon \cite{solomon}, and (b) the same model converted into a 3D stable structure using ChemDraw\textsuperscript{\textregistered}. Polar hydrogen atoms are colored in white, carbon in gray, nitrogen in blue, oxygen in red, and sulfur in yellow.}
	\label{fig:Cfig_Solomon_model}
\end{figure}

An initial model of coal was made. Starting with ChemDraw\textsuperscript{\textregistered}, 3D computer models of Pittsburgh No. 8 coal (hereafter referred to as P8 coal) were constructed from Solomon's proposed P8 coal, derived from structural and thermal decomposition data \cite{solomon}. P8 coal is a metallurgical, high-volatile (A) bituminous  ranked coal (hvAb) \cite{white1990}. Figure \ref{fig:Cfig_Solomon_model}a showcases the 2D representation of the P8 Solomon model, characterized by its chemical formula: C$_{161}$H$_{140}$O$_{15}$N$_{2}$S$_{2}$. 

The aliphatic and aromatic structures were initially delineated in 2D using ChemDraw\textsuperscript{\textregistered}. Subsequently, these structures were subjected to crude optimization through the MM3 force field \cite{MM3} to achieve a stable steric conformation (refer to Figure \ref{fig:Cfig_Solomon_model}b). The \texttt{PACKMOL} software package \cite{packmol} was used to construct supercell models, each comprising 5 units (1600 atoms) of the macromolecular unit structure. In the figures, hydrogen, carbon, nitrogen, oxygen, and sulfur atoms are visually distinguished by white, gray, blue, red, and yellow colors, respectively.

\begin{figure*}[t!]
	\centering
	\includegraphics[width=\linewidth]{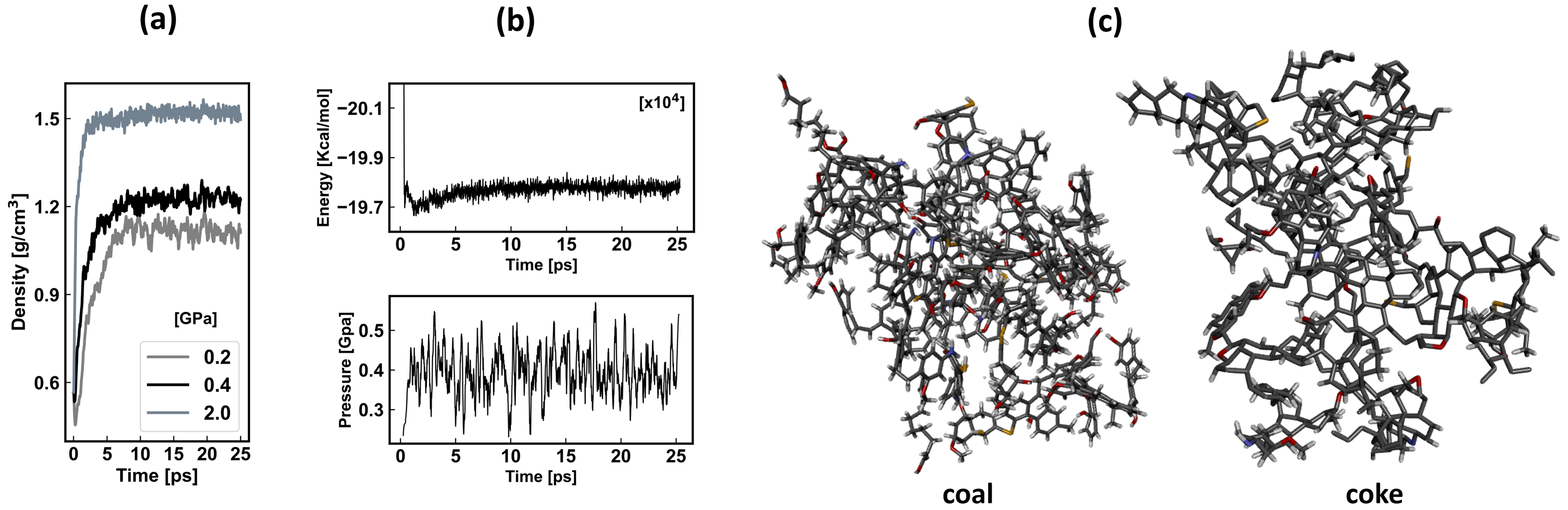}
	\caption{(a) The density time evolution of various pressures. The time development for the energy and pressure convergence at 0.4 GPa is shown in (b). The P8 coal model obtained from the NPT simulation shown (a) and (b), as well as the coke obtained after the NVT/NPT iterative simulation are also shown in (c).}
	\label{fig:Cfig_NPT}
\end{figure*}

\begin{table*}[b!]
	\caption{Details from the carbonization simulation at 1273 K. The chemical composition of the starting configuration is C$_{805}$H$_{700}$N$_{10}$O$_{75}$S$_{10}$.}
		\label{tab:detailsOFSim}
		\begin{tabular*}{\linewidth}{@{\extracolsep\fill}ccccc}

            \textbf{Pressure} & \textbf{P8 Coal Density} & \textbf{Coke Density} & \textbf{No. of Coke Molecules} &
            \textbf{Chemical Composition}  \\

             [GPa] & [g/cm$^3$] & [g/cm$^3$] &  &  \\

             \hline

             0.4 & 1.22 & 1.32 & 1 & C$_{611}$H$_{199}$N$_{5}$O$_{20}$S$_{4}$  \\
             
             2.0 & 1.49 & 1.66 & 1 & C$_{658}$H$_{200}$N$_{7}$O$_{10}$S$_{2}$ \\

	\hline
		\end{tabular*}
    
\end{table*}

The supercell models were extensively annealed using REAXFF to optimize their structure. New configurations were achieved through conjugate gradient relaxation, leading to local energy minima. To obtain realistic 3D coal models that reflect the density of P8 coal, the supercells were further annealed under isothermal-isobaric conditions (NPT), allowing density optimization. By doing so, the resulting models satisfy the periodic boundary conditions and therefore, can be used in accurate DFT calculations.

Vitrinite reflectance (R$_o$) is a well-established gauge of the interplay between temperature and pressure during coal maturation \cite{coalPressure}. Utilizing R$_o$ $=$ 0.78 for P8 coal \cite{Priv_Comm_vit} translates to a corresponding burial temperature of around 390 K \cite{VR1, VR3}. Consequently, the NPT simulation maintained a constant temperature of 400 K. Low pressure values (0.2, 0.4, and 2.0 GPa) were selected in alignment with experimental data from Reference \cite{coalPressure}, where a similar pressure range was used to establish the relationship between R$_o$ and pressure.

Figure \ref{fig:Cfig_NPT}a  shows the density time evolution for the simulations at 0.2, 0.4, and 2.0 GPa. Other relevant thermodynamic quantities were also continuously monitored to ensure simulation stability. The energy and pressure time-series plot at 0.4 GPa is depicted in Figure \ref{fig:Cfig_NPT}b. Models obtained at 2.0 GPa and 0.4 GPa were selected for this study due to their optimal densities ($\rho$) which align closely with experimental (1.46 g/cm$^3$) and estimated particle (1.22 g/cm$^3$) densities for Pittsburgh No. 8 coal, as reported by White and co-workers \cite{white1990}. The optimal densities obtained at 0.4 and 2.0 GPa are shown in the second column of Table \ref{tab:detailsOFSim}. 

The models were equilibrated at 500 K for 50 ps under constant temperature (NVT) conditions. Subsequently, the temperature was ramped to 1273 K over 100 ps. We introduce a new simulation protocol designed to capture the emission of volatiles and variations in density during coke formation in the carbonization process. This technique, referred to as "Simulation of Thermal Emission of Atoms and Molecules (STEAM)," is an iterative approach that incorporates successive NVT and NPT cycles. During the NVT phase, volatile molecules within a specified molecular mass range are systematically removed become non-bonded to the rest of the network. Meanwhile, the NPT phase ensures density convergence and maintains periodic boundary conditions in the system.

In our implementation of STEAM within this study, both the NPT and NVT phases extended over 125 ps each. Throughout the NVT phase, molecules with a molecular mass below 50 g/mol - formed due to bond cleavage - were removed at a rate of up to 5 molecules every 1.25 ps - modeling the outgassing. Completion of the STEAM process was determined by three criteria: (1) Absence of atom/molecule deletion in an NVT step, (2) predominance of high-molecular-mass molecules (preferably a substantial coke-forming molecule), and (3) maintenance of consistent pressure during NPT steps after the NVT phase in (1). Our simulation protocol is summarized as follows:

\begin{enumerate}
\item Formulate coal model and generate supercell of macromolecular coal units.
\item Determine initial model density via NPT simulation, leveraging R$_o$ data for the coal (if available).
\item Initiate STEAM implementation by selecting appropriate simulation duration for NVT and NPT phases. A recommended runtime should exceed 100 ps, ensuring sufficient time for density convergence during the NPT step.
\item Conduct NVT simulation, ensuring adequate atom removal intervals to allow a reasonable probability for molecule fragment recombination (e.g., start with 1000 timesteps).
\item Execute the NPT step while closely monitoring density convergence.
\item Iterate between NVT and NPT phases until no bond cleavage occurs in NVT step.
\item Re-run the last NVT step from \textit{item 6} to confirm absence of new bond cleavages.
\item Verify that mainly high-molecular-mass molecules remain, ideally a single molecule.
\item Perform an additional NPT step to attain a fully converged coke density.
\item Employ conjugate gradient relaxation to acquire an energy-minimized configuration.
\end{enumerate}

\noindent Figure \ref{fig:Cfig_NPT}c shows coal structure formed at 0.4 GPa ($\rho$ = 1.21 g/cm$^3$) after \textit{item 3} was completed, and the coke structure obtained after \textit{item 10}. A corresponding figure for the coal and coke formed at 2.0 GPa can be found in Figure S1 in the supplementary material.

\begin{figure*}[t!]
	\centering
	\includegraphics[width=\textwidth]{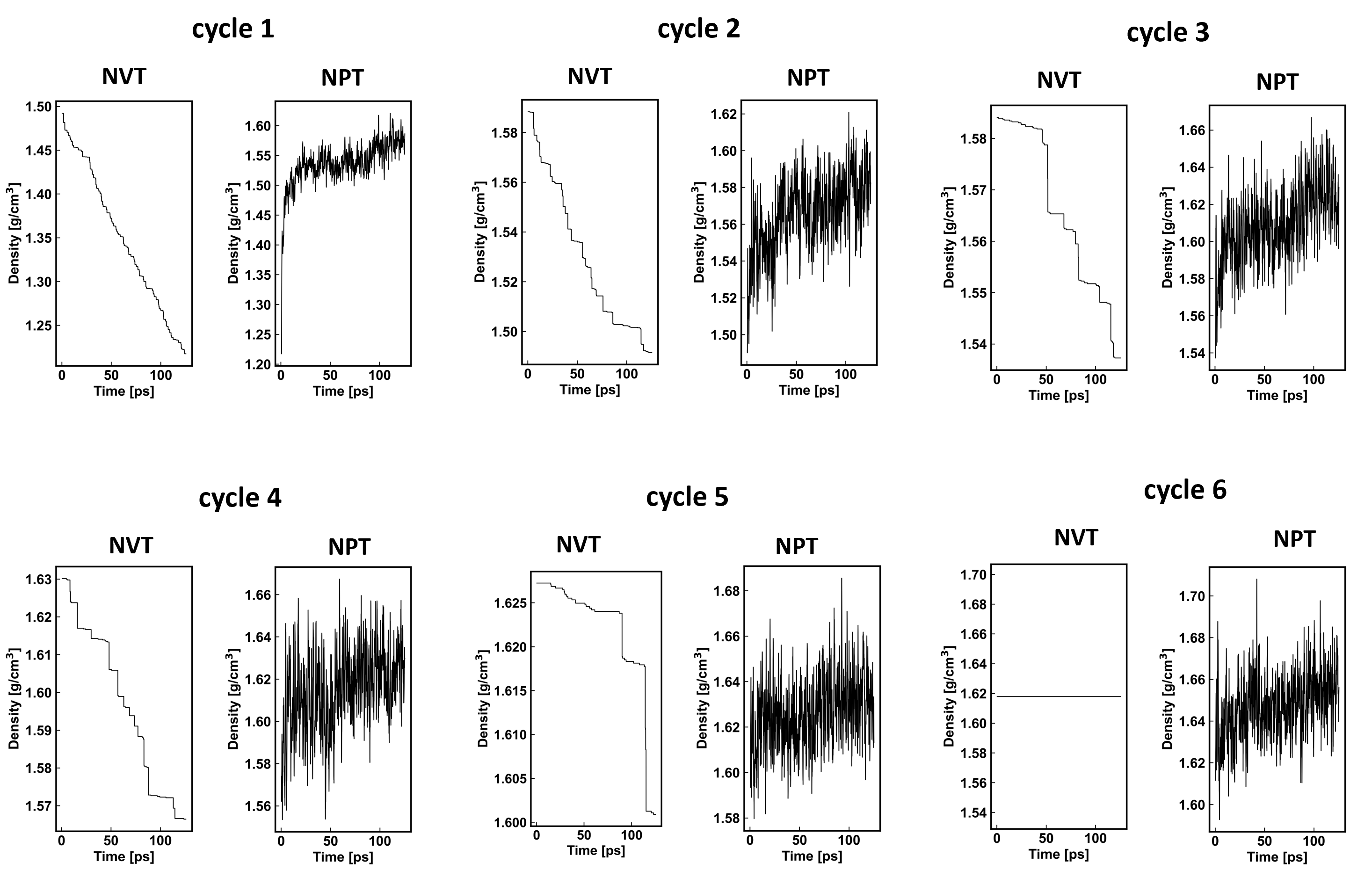}
	\caption{Iterative molecule removal process at 2.0 GPa. The NVT step in cycle 6 was repeated to ensure consistency. A similar plot obtained at 0.4 GPa is shown in the supplementary material.}
	\label{fig:Cfig_carbonizedData2p0}
\end{figure*}

\begin{figure*}[t!]
	\centering
	\includegraphics[width=\textwidth]{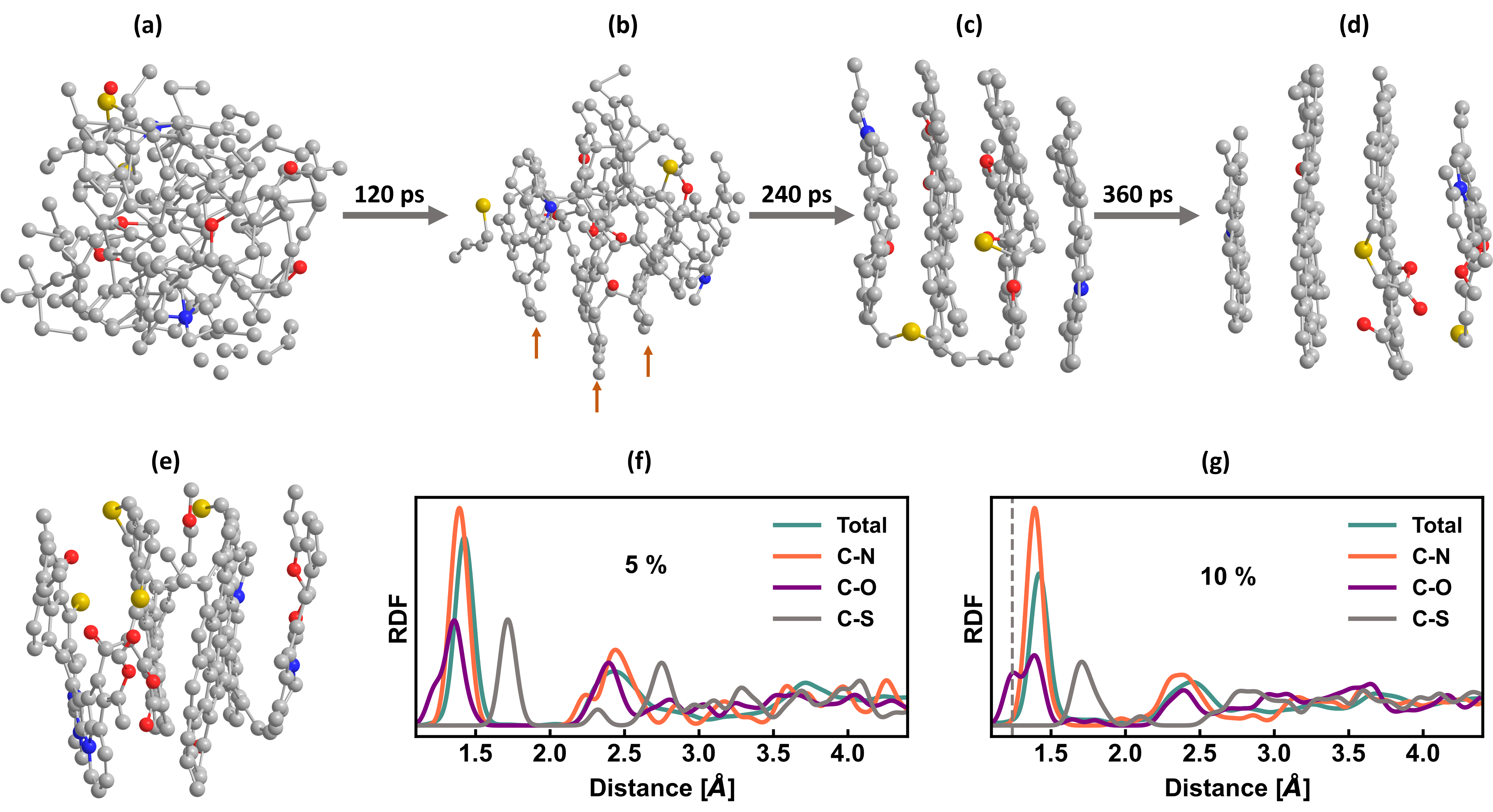}
	\caption{ (a - d) Chronology of formation of the layered nanostructure with 5\% impurity concentration included (using plane-wave DFT). The brown arrows in (b) pinpoint the putative planes. The forming structure from the model with 10\% impurity elements is displayed in (e). The corresponding radial distribution function obtained for impure amorphous graphite with (f) 5\% and (g) 10\% impurity concentrations.}
	\label{fig:Cfig_graphitization}
\end{figure*}

In the initial phases of carbonization, rapid emission of light gases was observed, succeeded by expulsion of heavier gases before the eventual formation of coke. This pattern is illustrated by the density-time plots for the NVT phase in the overview of the STEAM process at 2.0 GPa and 0.4 GPa, shown in Figure \ref{fig:Cfig_carbonizedData2p0} and Figure S2 respectively. Within the NVT step, the consistent volume scaling demonstrates that the release of low-mass molecules corresponds to minor step heights, while higher-mass molecules result in more prominent step heights. Noteworthy among the gases emitted are hydrogen (H$_2$, stemming from $\bullet$H radicals), carbon oxides (CO and CO$_2$), hydrogen sulfide (H$_2$S), hydrogen cyanide (HCN), water vapor (H$_2$O), as well as  hydrocarbons like CH$_4$,C$_2$H$_2$, C$_2$H$_4$, C$_2$H$_6$, and so on - gases intrinsically associated with the carbonization process. Occasionally, alkyl radicals (R$\bullet$), predominantly $\bullet$CH$_3$, combine with $\bullet$H radicals to yield alkanes.

Subsequent bond cleavage analysis revealed that the primary source of oxygen-related volatiles was hydroxyl (OH) present in carboxyl (R--COOH) or alcohol (R--OH) functional groups. In specific cases, ether (R-O-R$^\prime$) bond cleavage form formaldehyde (HCHO), which subsequently combined with $\bullet$OH radicals to produce CO and CO$_2$. As carbonization progressed, prior to coke formation, hydrogen sulfide (H$_2$S) emerge from mercaptans (R--SH). The delayed release of H$_2$S was also observed by Whittaker and Grindstaff \cite{irreversible}: The evolution of the sulfurous gas induces an internal pressure that results in the irreversible expansion of graphitic precursors in the late stages of carbonization, or the early stages of graphitization \cite{irreversible}. In rare instances, CS (with potential for evolution into CS$_2$) was liberated from heterocyclic thiophenes. Furthermore, bond fragmentation within aromatic rings resulted in aromatic hydrogen generation and triggered the liberation of hydrogen cyanide from ring nitrogen.

The compositions of the coke, as detailed in Table \ref{tab:detailsOFSim}, revealed that over 65\% of the non-carbon elements were emitted as gases, contrasting with the retention of about 80\% of carbon atoms. As depicted in Figure \ref{fig:Cfig_NPT}c and Figure S1, some carbon ring structures (5, 6, and 7-membered rings) were retained in the coke (indicative of the early stages of graphitization). The coke matrix retained pyrrolic and pyridinic nitrogen due to hydrogen atom removal from 5- and 6-membered heterocyclic aromatic rings, respectively. Aliphatic ethers (R--O--R$^\prime$) also persisted, serving as bridges between aromatic and aliphatic carbon structures. Cyclic ethers also formed as oxirane and oxolane structures. The Carbonyl components in P8 coal gave rise to emerging ketones (R$_2$C=O), and the Organosulfur compounds manifested as thioethers (R--S--R$^\prime$) and thioketones (R$_2$C=S).

\subsection{Graphitization \label{sec:vasp}}

To study graphitization, we found that utilizing DFT precision forces is required \cite{aG, aG2}; thus, in this section we employed \texttt{VASP}. Our approach involves randomly distributing 200 atoms of carbon, nitrogen, oxygen, and sulfur in a cubic box to achieve a density of 2.4 g/cm$^3$. Henceforth the non-carbon elements will be referred to as ``impurity elements". Six models were created, incorporating 5\% and 10\% impurity elements (three models for each concentration). The oxygen to nitrogen and oxygen to sulfur ratio was 3:1 for both cases. This ratio was deliberately selected to reflect the elemental concentration in the coke (as detailed in Table \ref{tab:detailsOFSim}), while maintaining a carbon-rich environment. The models were annealed for 360 ps at 3000 K - the common graphitization temperature \cite{acheson, coalToGraphiteTemp1,coalToGraphiteTemp2}. Subsequently, the models were relaxed to an energy minimum configuration using the conjugate gradient algorithm. Note that hydrogen is excluded. This is because hydrogen is typically ``burned off" in the carbonization phase, and frankly, the timestep required is too short for practical \texttt{VASP} simulations.

The interactions between the electrons and ions were described using the Perdew-Burke-Ernzerhof projected-augmented-wave pseudopotential \cite{PBE, PAW}. We set a cutoff energy of 400 eV for the plane-wave-basis used to expand electronic wave functions during the molecular dynamics simulation, and a higher cutoff of 520 eV was used for the electronic structure calculations. For where it applies, the layered nanostructures containing non-carbon elements are referred to as impure amorphous graphite.

Figure \ref{fig:Cfig_graphitization}a - d is a chronological depiction of the layering process observed in layered nanostructure with 5\% impurity concentration. In contrast, the models with 10\% impurity concentration lack distinct layers; instead, they exhibit inter-layer atomic bonding involving carbon atoms near oxygen atoms (see Figure \ref{fig:Cfig_graphitization}e). Wang and Hu highlighted the connection between layering defects in graphite and the presence of oxygen, revealing that even at low concentrations, graphite oxidation significantly disrupts its layered structure \cite{GO2}.

Preceding any discernible indications of layering (indicated by the brown arrows in Figure \ref{fig:Cfig_graphitization}b), a process involving rearrangement of the non-carbon atoms occurred. Nitrogen atoms readily adopt graphitic nitrogen (N-3) structures (forming bonds with three neighboring carbons in sp$^2$ configuration). Once formed, this configuration was maintained throughout the simulation. In specific cases, nitrogen atoms disrupt ring connectivity in the layers by bonding with only 2 carbon atoms after substituting a sp$^2$ carbon atom, thereby forming pyrrolic (N-5) or pyridinic (N-6) structure by carbon substitution in a 5- or 6-membered ring, respectively. Oxygen atom form ether-bridges (C--O--C) or ketone (C=O) structures that terminate ring connectivity. Similarly, sulfur atoms display a preference for thioether (C--S--C) and thioketone (C=S) structures. Notably, sulfur conformation in impure amorphous graphite mirrors sulfur's inclimation to form C--S--C bond at the edges (grain boundaries) in crystalline graphite \cite{adjizian2013dft}.

In early graphitization stages, sulfur atoms initially establish crosslinks between layers (see Figure \ref{fig:Cfig_graphitization}c). However, upon energy optimization, sulfur atoms demonstrate a preference for bonding within the layers (refer to Figure \ref{fig:Cfig_graphitization}d). From experimental observations, Kipling and co-workers suggested that crosslinking of sulfur between layers exist in graphitized materials from sulfur-containing precursors \cite{sulfurCross} This was later challenged by Adjizian and co-workers in a DFT study that found such sulfuric crosslinking structures were energetically unfavorable in graphite \cite{adjizian2013dft}. Our results offer a nuanced perspective: crosslinking does occur during intermediate graphitization stages, but the energy-optimized sulfur conformation is not as a crosslink between the layers. Additionally, our observation from the structures with 10\% non-carbon elements also contradicts Kipling's notion that higher sulfur concentrations could promote inter-layer connections; instead, distinct layers simply failed to form, atleast on the timescale of our simulation.

The functional groups in impure amorphous graphite were also identified in coke post-carbonization (see Section \ref{sec:reaxff}), indicative of their energy stability within carbon layers during coal graphitization \cite{landmark1}. This notion is reinforced by NMR analysis of graphite oxide, revealing stable ether (C--O--C) structures and unstable hydroxyl (C--OH) groups, with the transient C--OH condensing into consolidated C--O--C linkages \cite{GO}. Furthermore, a predominant sulfur-doped graphite conformation reported by Li \textit{et al.} is the theophinic (C--S--C) structure \cite{GS}.

We analyzed the local structure in impure amorphous graphite using the radial distribution function (Figure \ref{fig:Cfig_graphitization}f and g). Key details about the primary peak, which represents the nearest distance between carbon and non-carbon pairs, are provided in Table \ref{tab:bondlenght}. Our analysis is limited to this range due to the low impurity concentration in the models, which inherently lacks extensive structural information beyond immediate neighboring atoms. Notably, the nearest neighbor values derived from our models closely correlate with experimental data for the observed functional groups, as discussed earlier. Another feature is the emergence of a distinct peak around 1.24 $\AA$ in the radial distribution function plot of the 5\% impurity model (Figure \ref{fig:Cfig_graphitization}f), corresponding to the C=O bond length in carbonyl groups \cite{carbonyl}. This peak becomes more prominent in the 10\% impurity model's plot (highlighted by gray dashed lines in Figure \ref{fig:Cfig_graphitization}g), suggesting a higher oxygen concentration encourages the formation of carbonyl functional groups in impure amorphous graphite.

\begin{table}[h!]
	\caption{Nearest neighbor bond length of the models compared to experimental data on bond length from other works. The unit of measurement is in $\AA$.}
		\label{tab:bondlenght}
		\begin{tabular*}{\linewidth}{@{\extracolsep\fill}ccccc}

                
                
			  \textbf{C--C} & \textbf{C--N} & \textbf{C--O} & \textbf{C=O} & \textbf{C--S}\\
			\hline
			1.42 \AA  & 1.40 \AA  & 1.24 \AA & 1.38 \AA & 1.71 \AA  \\
			  1.42 \cite{graphite} & 1.47 \cite{graphiticN} & 1.23 \cite{carbonyl} & 1.43 \cite{ether} & 1.73 \cite{thioether} \\
	\hline
		\end{tabular*}
\end{table}

\begin{figure}[b!]
	\centering
	\includegraphics[width=\linewidth]{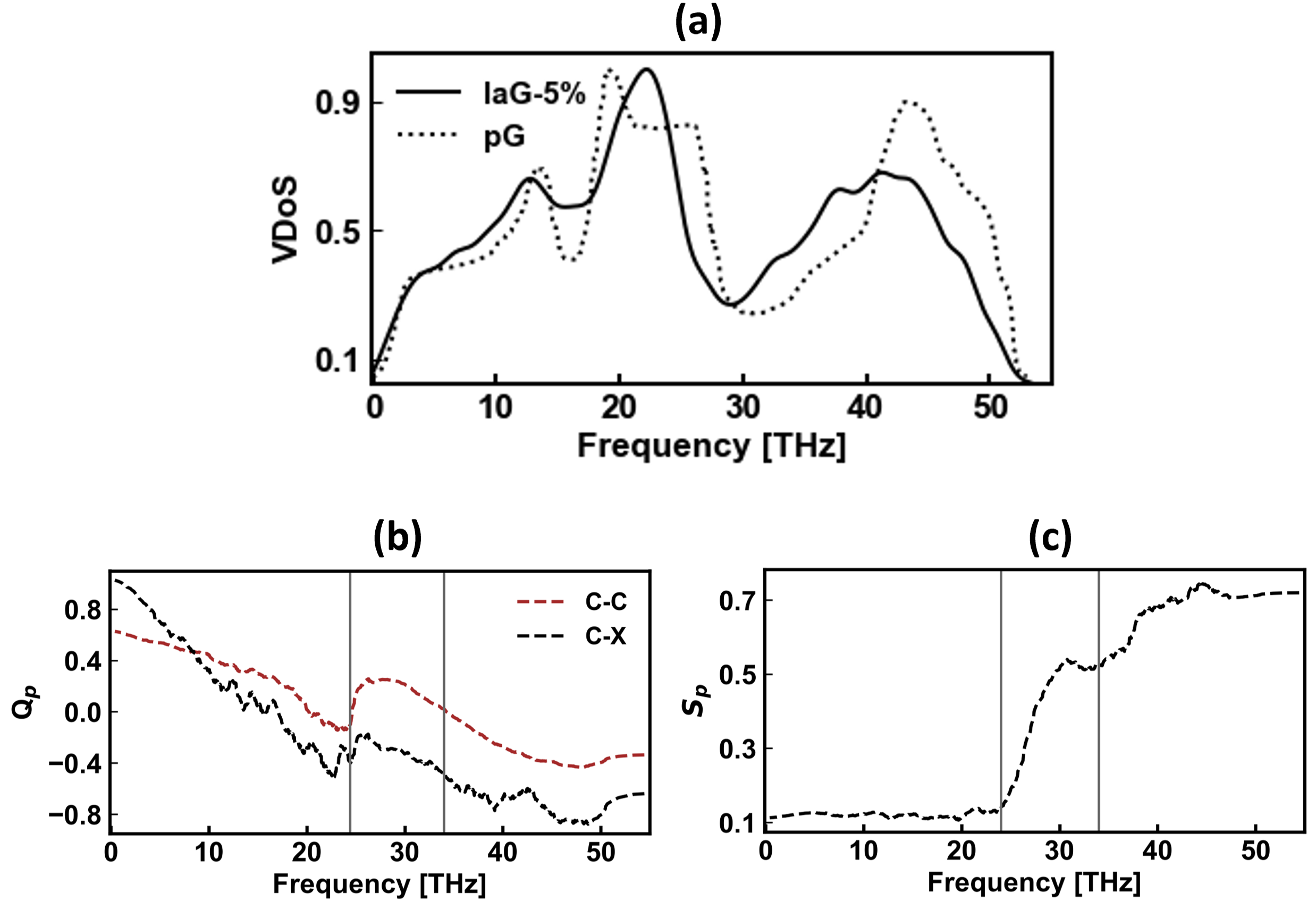}
	\caption{(a) Vibrational density of states (VDoS) for impure amorphous graphite with 5\% non-carbon elements (IaG-5\%) compared with that of pristine graphite (pG). The phase quotient (Q$_p$) describing the vibrational modes between carbon atoms (C) and between carbon and non-carbon atoms (X) is shown in (b). The bond stretching and bending (S$_p$) in amorphous layered structure is shown in (c).}
	\label{fig:Cfig_VDoS}
\end{figure}

The vibrational signatures in impure amorphous graphite were extracted and compared with that of pristine graphite \cite{aG} by computing the vibrational density of states (VDoS) using the harmonic approximation. This entailed calculating the Hessian matrix by evaluating forces from atomic displacements of 0.015 $\AA$ along six directions ($\pm$x, $\pm$y, $\pm$z). The VDoS ($g(\omega)$) is computed as follows:

\begin{equation}
    g(\omega) = \frac{1}{3N} \sum_{i=1}^{3N} \delta(\omega - \omega_i) 
    \label{eqn:VDoS}
\end{equation}

\noindent where $\delta(\omega - \omega_i)$ is approximated as a Gaussian with a standard deviation of 1.5\% of the maximum frequency observed. $N$ and $\omega_i$ represent the number of atoms and the eigen-frequencies of normal modes, respectively. The VDoS for the pristine and impure amorphous graphite have similar structures but with shifts in peak location (refer to Figure \ref{fig:Cfig_VDoS}a). A shoulder appearing at the low-frequency end (around 5 THz) matched in both models. Moreover, the peak with a shoulder around 21 THz in the pristine graphite model was resolved as a single peak in impure amorphous graphite.

While there are noticeable similarities in the vibrational structures between pristine graphite and impure amorphous graphite, it's important to recognize that the classification of vibration modes applied to pristine graphite, including acoustic and optical modes, cannot be directly extended to describe amorphous systems. To address this disparity, we rely on the concept of the phase quotient (Q$_p$) \cite{Bell_1975}, which acts as a metric to discern whether vibrations can be characterized as acoustic (+Q$_p$) or optical (-Q$_p$) modes. The phase quotient is derived as  \cite{SiOx}:

\begin{equation}
    Q_p ~= \frac{1}{N_b}\frac{\sum_{m} \boldsymbol{u}^{i}_{p} \cdot \boldsymbol{u}^{j}_{p}}{\sum_{m}| \boldsymbol{u}^{i}_{p} \cdot \boldsymbol{u}^{j}_{p}|}  
    \label{eqn:Qp}
\end{equation}

\noindent where $N_b$ is the number of valance bonds, and $\boldsymbol{u}^{i}_{p}$ and  $\boldsymbol{u}^{j}_{p}$ are the normalized displacement vectors for the $p^{th}$ normal mode. The index, $i$, ranges across all carbon atoms, and $j$ enumerates neighboring atoms (C, O, N, S) linked to the $i^{th}$ carbon atom. A purely acoustic (optical) vibration manifests as a phase quotient of +1 (-1). 

In Figure \ref{fig:Cfig_VDoS}b, Q$_p$ profiles are illustrated as crimson and black curves, representing carbon/carbon (C--C) and carbon/non-carbon (C--X) vibrations. Particularly noteworthy is the stronger in-phase (acoustic mode) coherence of C--X vibrations at very low frequencies, in contrast to C--C vibrations. This coherence in C--X vibrations transitions rapidly to an out-of-phase pattern — akin to optical modes — even surpassing the out-of-phase behavior of C--C vibrations at higher frequencies.

\begin{figure*}[b!]
	\centering
	\includegraphics[width=\linewidth]{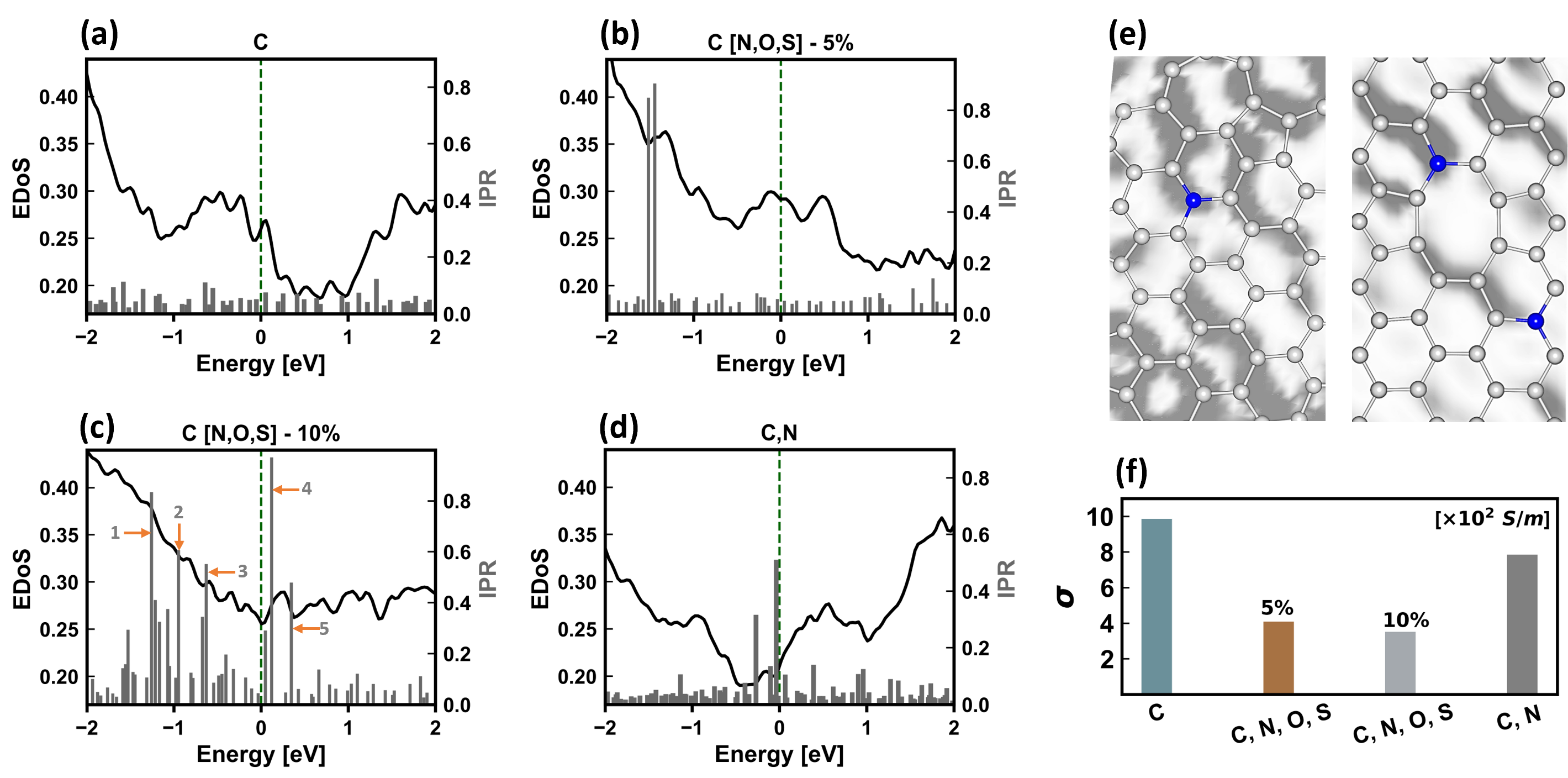}
	\caption{Electronic density of states (EDoS) calculated for (a) amorphous graphite (only carbon atoms present), impure amorphous graphite with (b) 5\% and (c) 10 \% impurity concentration, and (d) 5\% nitrogen impurities only. In (a) - (d), the Fermi level (E$_f$) is shifted to zero and indicated by green dashed lines, and only the region between E$_f$ $\pm$ 2 eV is shown. The complete EDoS is provided in Figure S3. SPC calculation \cite{SPC1}, presented as a grayscale heatmap on selected layers containing nitrogen, is shown in (e); the darker regions represent the electronic conduction paths in the layered nanostructures. The average electronic conductivity ($\sigma$, in S/m) for all 3 models is plotted in (f).}
	\label{fig:Cfig_EDoS_SPC}
\end{figure*}

In graphite, the acoustic mode is commonly linked with the low-frequency region $<$ 26.8 THz \cite{C5CP03466C}. The Q$_p$ for C--C vibration in impure amorphous graphite shifted to the optical mode (-Q$_p$) at $\approx$ 21 THz. However, a local minimum emerged at Q$_p$ = -0.15, corresponding to a frequency of 24.4 THz (indicated by the first black line in Figure \ref{fig:Cfig_VDoS}b). The vibration mode briefly reverted to the acoustic mode, peaking at 27 THz (the vibration mode-switching frequency in graphite), before Q$_p$ transitioned back towards the optical mode. Q$_p$ became negative (optical mode) at $\approx$ 34 THz, indicated by the second black line in Figure \ref{fig:Cfig_VDoS}b. Two potential explanations underlie this behavior. Firstly, the turning point of the phase quotient for C--X, occuring at a lower frequency of 22.6 THz, may have influenced the C--C vibrational mode, inducing a switch from optical to acoustic vibrational behavior. Secondly, the region experiencing a shift in slope within the phase quotient plot corresponds to the transition from bending to stretching vibrational characteristics. This transition is evaluated using the bending/stretching quotient (S$_p$) proposed by Marinov and Zotov \cite{marinov1997model}:

\begin{equation}
    S_p ~= \frac{\sum_{m} |(\boldsymbol{u}^{i}_{p} - \boldsymbol{u}^{j}_{p}) \cdot \hat{\boldsymbol{r}}_{ij}|}{\sum_{m}| \boldsymbol{u}^{i}_{p} - \boldsymbol{u}^{j}_{p}|}  
    \label{eqn:Sp}
\end{equation}

\noindent Here, $\boldsymbol{u}^{i}_{p}$ and $\boldsymbol{u}^{i}_{p}$ are the same as in Equation \ref{eqn:Qp}, and $\hat{\boldsymbol{r}}_{ij}$ is the unit vector parallel to the m$^{th}$ bond. S$_p$ $\to$\ 0 indicates bond-bending character, and S$_p$ $\to$\ 1 represent predominantly bond-stretching vibrations. The region between 24.4 - 34 THz, marked by two black vertical lines in Figure \ref{fig:Cfig_VDoS}c, signifies a transition from bending to stretching vibrational character. Significantly, this transition aligns with the region of C--C vibration mode switching, as indicated by Q$_p$ (see Figure \ref{fig:Cfig_VDoS}b), implying a direct influence of vibrational character transition on vibrational modes in impure amorphous graphite.

In Figure \ref{fig:Cfig_EDoS_SPC}a-c, the electronic density of states (EDoS) is provided for three configurations: (a) amorphous graphite, and impure amorphous graphite with (b) 5\% and (c) 10\% impurity concentrations. The Fermi level (E$_f$) in the plots have been shifted to zero (indicated by the green dashed line), and only regions, E$_f$ $\pm$2 eV, are displayed here. The complete electronic structures, plotted in Figure S3a, indicate minimal alterations to the overall electronic structure of amorphous graphite, when compared to the impure amorphous graphite in Figure S3b and c for 5\% and 10\% impurity concentrations. We use the inverse participation ration (IPR) to study localization of the electronic states. The IPR ($\zeta_n$) is defined as:

\begin{equation}
\zeta_n = {\sum_i {| \gamma_n^i |^4 } \over (\sum_i {| \gamma_n^i |^2 } )^2}
\label{eqn:IPR}
\end{equation}

\noindent where $\gamma_n^i$ represents the contribution of the i$^{th}$ Kohn-Sham state to a given eigenvector ($\epsilon_n$). A high IPR value ($\zeta_n$ $\to$\ 1) indicates that the wave function is localized on very few atoms. Conversely, a low value ( $\zeta_n$ $\to$\ 0) indicates that the wave function is delocalized (distributed over many atoms).

The gray dashed lines in Figure \ref{fig:Cfig_EDoS_SPC}a - c represent the IPR for the electronic bands. In amorphous graphite, the low IPR suggests the absence of localized states near E$_f$. However, the introduction of impurity elements leads to localized states near E$_f$ (see Figure \ref{fig:Cfig_EDoS_SPC}b and c). 

We identified the elemental species responsible for the highly localized states in impure amorphous graphite, focusing on IPR values $\gtrsim$ 0.5. In the layered nanostructure with 5\% impurity concentration, two such states emerge, separated by 0.07 eV, and they are exclusively associated with a single ketone oxygen atom. When the impurity concentration is increased to 10\%, additional localized states surface. In Figure \ref{fig:Cfig_EDoS_SPC}c, states 1, 2, and 3 (marked by brown arrows) localize on distinct oxygen atoms forming ketones, while state 4 localizes on a thioketone. State 5, with IPR $\approx$ 0.48, is also centered on the same thioketone as state 4, albeit separated by 0.2 eV

We note that the nitrogen atoms in impure amorphous graphite do not contribute to the localized states around the Fermi energy ($E_f$). Extensive research has focused on intentionally doping nitrogen into various carbon allotropes \cite{stumm1995structural, stumm1997defects}. For example, nitrogen has been introduced into graphite (or graphene) to create nitrogen-doped quantum dots \cite{Qdot1}, and into diamond to form nitrogen-vacancy centers for potential super-computing applications \cite{NVC}. With this context, we computed the electronic structure for impure amorphous graphite containing 5 weight-percent of only nitrogen atoms, as shown in Figure \ref{fig:Cfig_EDoS_SPC}d and Figure S3d. The models were constructed by repeating the simulation protocol detailed in Section \ref{sec:vasp}, and a representation of the nitrogen-containing impure amorphous graphite model is provided in Figure S3e.

As shown in Figure S3d, the overall electronic structure of the layered structure with 5\% nitrogen, exhibits minimal differences when compared to that of amorphous graphite (see Figure S3a). Notably, two states emerge in proximity to the Fermi level (refer to Figure \ref{fig:Cfig_EDoS_SPC}d). However, these states (with IPR values of 0.32 and 0.51) do not exclusively localize on specific atoms; instead, they are distributed across a small number of atoms, implying an absence of true localization. This observation is further illustrated in Figure S3f for the state with the higher IPR value (IPR = 0.51). 

Interestingly, while nitrogen atoms substitute for carbon atoms within the sp$^2$ configuration of the layers, they do not contribute to localizing electronic states near $E_f$. This leads to the question:  Does nitrogen merely emulate carbon behavior within the layers, and how does it impact electronic conductivity in this layered nanostructure? To address this, we compute the electronic conduction and transport in the nitrogen-containing impure amorphous graphite, utilizing the space-projected conductivity (SPC) methodology \cite{SPC1}.

The SPC framework utilizes the Kubo-Greenwood formula for electronic conductivity \cite{Kubo,Greenwood} and the Kohn-Sham single-particle states ($\psi_{i,\textbf{k}}$) to project the electronic conductivity from a given atomic arrangement onto a spatial grid (see Reference \cite{SPC1} for details on the SPC method). Figure \ref{fig:Cfig_EDoS_SPC}e depicts the electronic transport path using a grayscale heatmap within selected nitrogen-containing regions. Darker shades indicate high electronic conduction, while lighter areas indicate limited or absent electronic transport. Notably, the presence of nitrogen atoms along a conduction pathway acts to impede or completely halt electronic transport. To examine the potential impact of structural attributes on electronic conductivity originating from nitrogen within the layered nanostructure, we computed the radial distribution function (RDF) and bond angle distribution (BAD) for nitrogen-containing impure amorphous graphite (Figure S4a and b, respectively). The most prominent RDF peak, representing the average nearest neighbor distance calculated for carbon-carbon (C--C) and carbon-nitrogen (C--N) bonds, were closely aligned - separated by only 0.015 $\AA$. The bond angle distribution (BAD) for central carbon (C--C--C) and central nitrogen (C--N--C), as depicted in Figure S4b, displayed similar pattern, with both having a maximum peak around 118$\circ$ (close to the 120$\circ$ angle in graphite). These suggest that nitrogen substitutions within the layered nanostructure maintain the precise sp$^2$ layered arrangement of the carbon atoms. Consequently, the impedance of electronic conduction pathways by nitrogen atoms does not stem from their structural attributes within impure amorphous graphite. 

Our prior research highlighted that electronic conductivity in amorphous graphite favors path with interconnected 6-membered (hexagonal) rings \cite{aG}. Building on this understanding, we now observe that the introduction of nitrogen atoms, even within a 6-membered ring arrangement, disrupts electronic conduction within the planes. This insight suggests that purposefully incorporating nitrogen into graphite derived from other carbonaceous materials could provide unique avenues for electronic conduction. The average electrical conductivity ($\sigma$) was calculated for both amorphous graphite and impure amorphous graphite with varying impurity weight percentages (see Figure \ref{fig:Cfig_EDoS_SPC}f). The electronic conductivity of pure carbon amorphous structure measured significantly lower by a factor of 100 compared to crystalline graphite \cite{aG}. The electronic conductivity of nitrogen-containing impure amorphous graphite was $\approx$ 20\% lower than that of amorphous graphite. The incorporation of oxygen and sulfur impurities further decreased the conductivity values. In comparison to amorphous graphite, the electronic conductivity dropped by 58\% and 64\% in impure amorphous graphite with 5\% and 10\% impurity concentrations, respectively. These estimates are qualitative but indicative of the transport consequences of the impurities.

\section{Conclusions}

The thrust of this work is to offer a first-of-its-kind, pragmatic perspective into the structure, vibrational behavior, and electronic properties of impure amorphous graphite, derived from coal-like atomistic models through high-temperature transformation.Employing a novel simulation protocol with the REAXFF potential, designed to replicate the initial carbonization stages leading to single-molecule coke formation, this research unveils enduring functional groups containing non-carbon elements during the pyrolysis of bituminous coal at 1200 K. Leveraging \textit{ab initio} DFT, the graphitization of models featuring coal-like elemental compositions, including carbon atoms with 5 and 10 weight-percent of nitrogen, oxygen, and sulfur atoms, is explored. The findings emphasize the overall vibrational and electronic structure of the impure amorphous graphite bear significant similarity to that of amorphous graphite. However, the presence of nitrogen, particularly in graphitic form, impedes electronic conductivity within layers. This research advances both methodologies and insights for exploring coal's alternative applications beyond energy production, offering valuable contributions to its integration into cutting-edge electronic technologies.

\section*{Acknowledgments}
\noindent C.U. extends gratitude to Mr. Dan Connell and to CONSOL Energy Inc. for providing the invaluable research internship opportunity (at CONSOL Innovations LLC) which played a pivotal role in facilitating the successful culmination of this work. C.U. also express his appreciation to the Nanoscale \& Quantum Phenomena Institute (NQPI) for the financial support conferred through the NQPI research fellowship. The Authors thank Ms. Anna-Theresa Kirchtag for proofreading the manuscript.\\

\noindent This material is based upon work supported by the Department of Energy under Award Number DE-FE0032143. It also used computational resources at Pittsburgh Supercomputing Center (Bridges-2 Regular Memory) through allocations phy230007p and dmr190008p from the Advanced Cyberinfrastructure Coordination Ecosystem: Services \& Support (ACCESS), supported by National Science Foundation grants; 2138259, 2138286, 2138307, 2137603, and 2138296.


\bibliographystyle{elsarticle-num} 
\bibliography{aGN}

\end{document}


\begin{frontmatter}


\title{Supplementary material- Computer Simulation of Carbonization and Graphitization of Coal}

\author[OU]{C. Ugwumadu \corref{cor} \fnref{lb2}}
\ead{cu884120@ohio.edu}
\author[CONSOL]{R. Olson III}
\author[CONSOL]{N. L. Smith}
\author[OU]{K. Nepal}
\author[RUSS]{Y. Al-Majali}
\author[RUSS]{J. Trembly}
\author[OU]{D. A. Drabold \corref{cor}}
\ead{drabold@ohio.edu}

\address[OU]{Department of Physics and Astronomy,
Nanoscale and Quantum Phenomena Institute (NQPI)\\
Ohio University, Athens, OH 45701, USA}

\address[CONSOL]{CONSOL Innovations, Triadelphia, WV 26059, USA}

\address[RUSS]{Department of Mechanical Engineering,
Institute for Sustainable Energy and the Environment (ISEE), \\
Ohio University, Athens, Ohio 45701, USA.}

\cortext[cor]{Corresponding Authors}

\fntext[lb2]{C.U. served as a Research Intern at CONSOL Innovations LLC during the preparation of this manuscript.}

\begin{abstract}
This study describes computer simulations of carbonization and graphite formation, including the effects of hydrogen, nitrogen,
oxygen, and sulfur. We introduce a novel technique to simulate carbonization, ``Simulation of Thermal Emission of Atoms
and Molecules (STEAM),” designed to elucidate the removal of volatiles and density variations in carbonization residue. The
investigation extensively analyzes the functional groups that endure through high-temperature carbonization and examines the
graphitization processes in carbon-rich materials containing non-carbon “impurity elements”. The physical, vibrational, and
electronic attributes of impure amorphous graphite are analyzed, and the impact of nitrogen on electronic conduction is investigated,
revealing its substitutional integration into the sp$^2$ layered network.
\end{abstract}

\begin{keyword}
coal, carbon, amorphous graphite, molecular simulation, carbonization, graphitization
\end{keyword}

\end{frontmatter}



\section{Overview}

\begin{figure*}[t!]
	\centering
	\includegraphics[width=\textwidth]{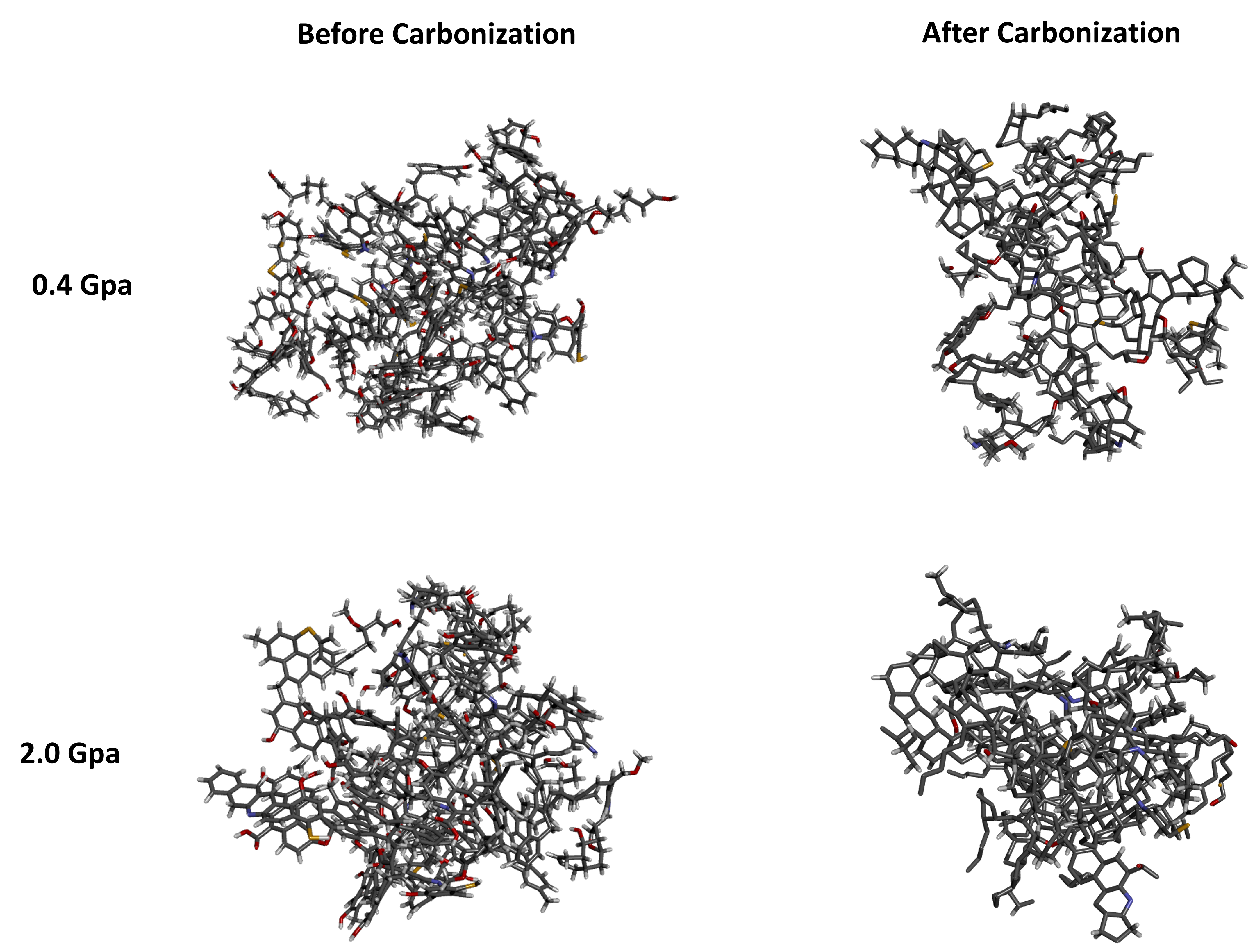}
	\caption{The starting and final configurations from both the 0.4 GPa and 2.0 GPa simulations are displayed. Following the process of carbonization, the ultimate configuration reveals the presence of five- and six-membered rings alongside sp carbon chains. A small number of hydrogen, nitrogen, oxygen, and sulfur atoms are retained in coke. The hydrogen, carbon, nitrogen, oxygen, and sulfur atoms are colored white, gray, blue, red, and yellow, respectively.}
	\label{fig:CSfig_carbonizaton}
\end{figure*}

\begin{figure*}[t!]
	\centering
	\includegraphics[width=\textwidth]{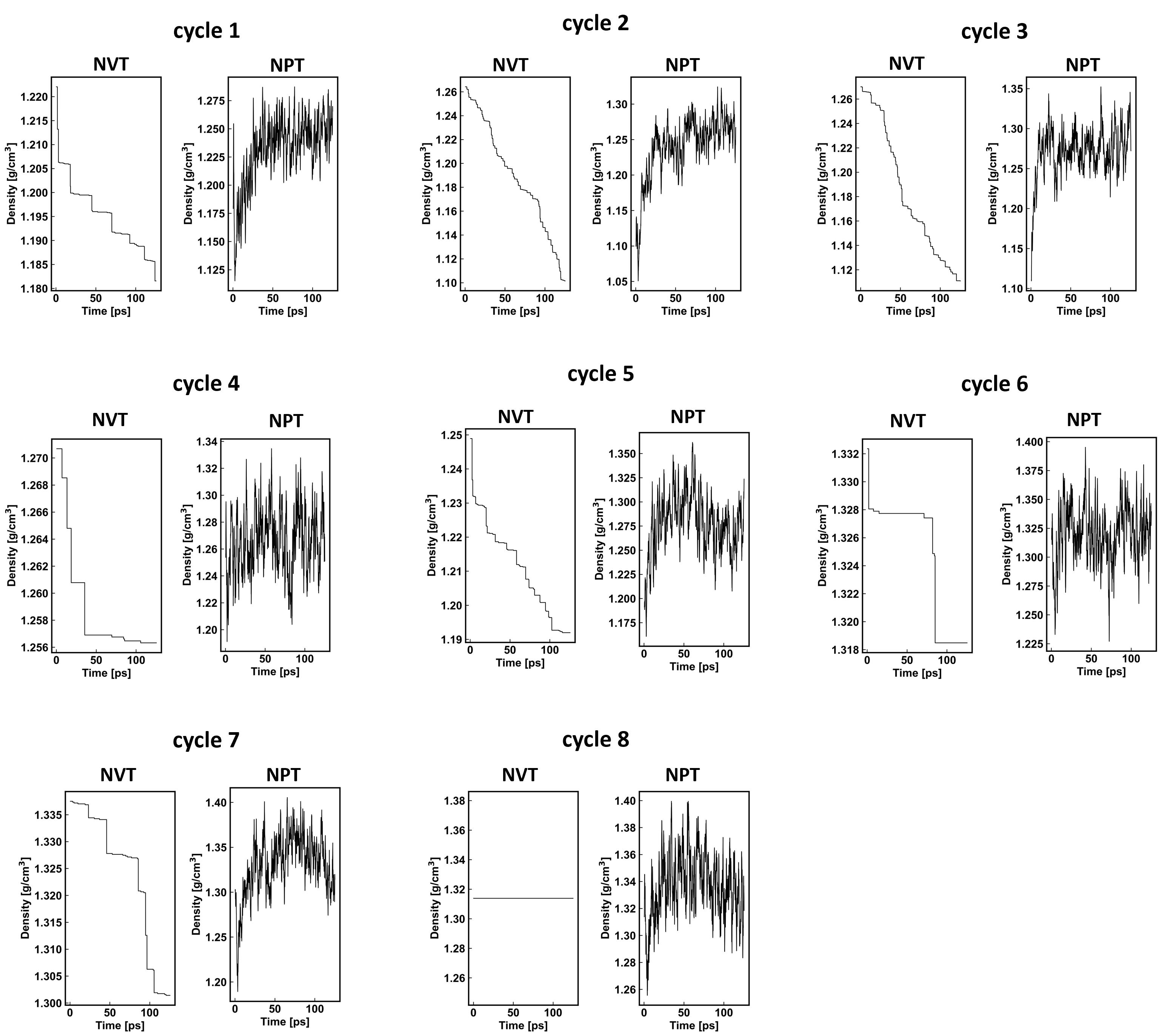}
	\caption{An overview of the simulation cycle illustrating the formation of coke under a pressure of 0.4 GPa is shown. The main manuscript elaborates on a detailed, systematic methodology that was employed to achieve these results. Notably, in the eighth cycle, we took the precautionary step of reiterating the final NVT phase. This measure was taken to ensure that no new bond cleavage events could occur during this stage of the simulation. The coke formed from this particular simulation is also presented in Figure \ref{fig:CSfig_carbonizaton}.}
	\label{fig:CSfig_carbonizedData0p4}
\end{figure*}

We include additional figures to supplement our main discussion. Figure \ref{fig:CSfig_carbonizaton} displays the initial P8 coal configuration post-density optimization at 0.4 GPa and 2.0 GPa, along with resulting coke models after STEAM implementation. These models retain 5- and 6-membered rings with residual non-carbon elements. Figure \ref{fig:CSfig_carbonizedData0p4} outlines the STEAM process for 0.4 GPa coke, while a similar plot for 2.0 GPa coke is in the manuscript.

\begin{figure}[t!]
	\centering
	\includegraphics[width=\textwidth]{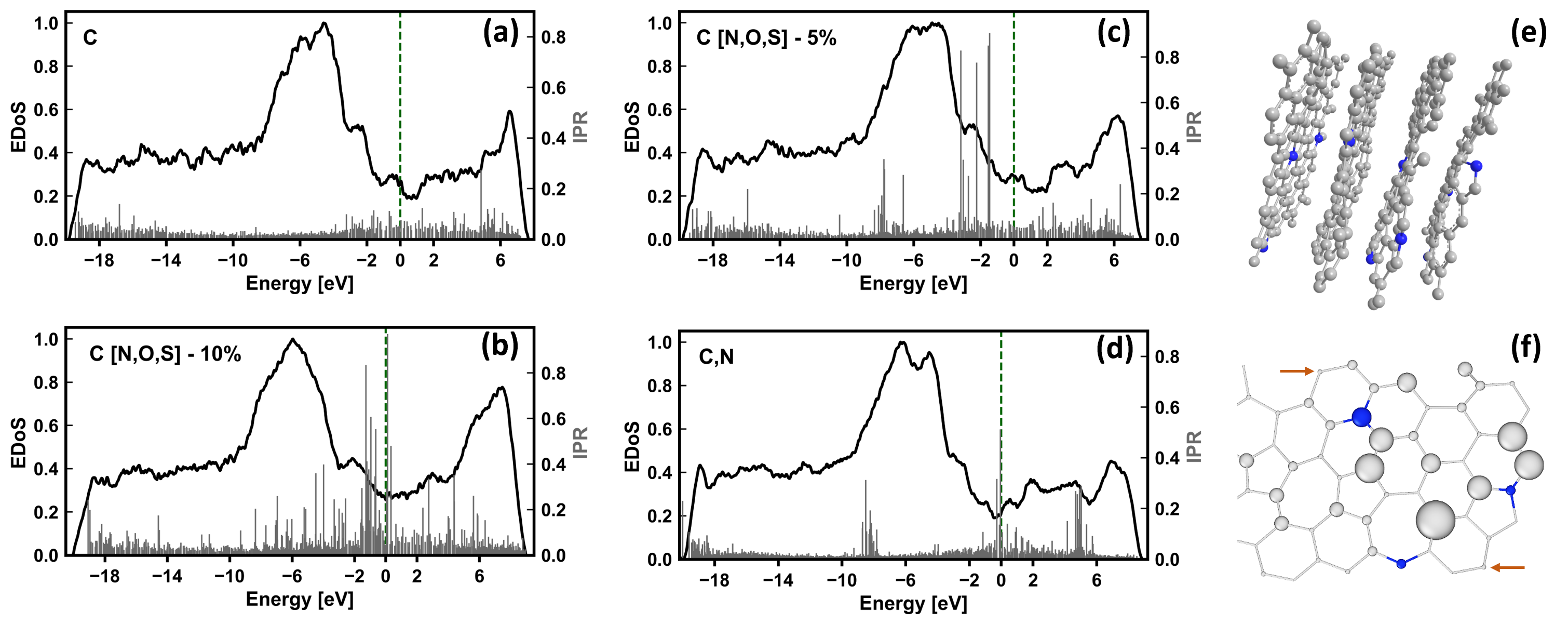}
	\caption{The electronic density of states (EDoS) for pure amorphous graphite (containing only C atoms) is shown in (a). The EDoS for impure system is displayed for   (b) 5\% impurity concentration, (c) 10\% impurity concentration, and (d) only 5\% nitrogen impurities. The green dashed lines indicate the shifted Fermi level, set at zero for all cases (a - d). Despite the introduction of impurities, the fundamental electronic structure of amorphous graphite remains largely unaltered. A the model of amorphous graphite with a 5\% nitrogen impurity concentration is depicted in (e). In (f), the distribution of localized states in this model (in e) is portrayed. The relative size of the atoms visually represents the degree of localization, with non-localized atoms indicated by brown arrows.}
	\label{fig:CSfig_FullEDoS}
\end{figure}

\begin{figure*}[h!]
	\centering
	\includegraphics[width=\textwidth]{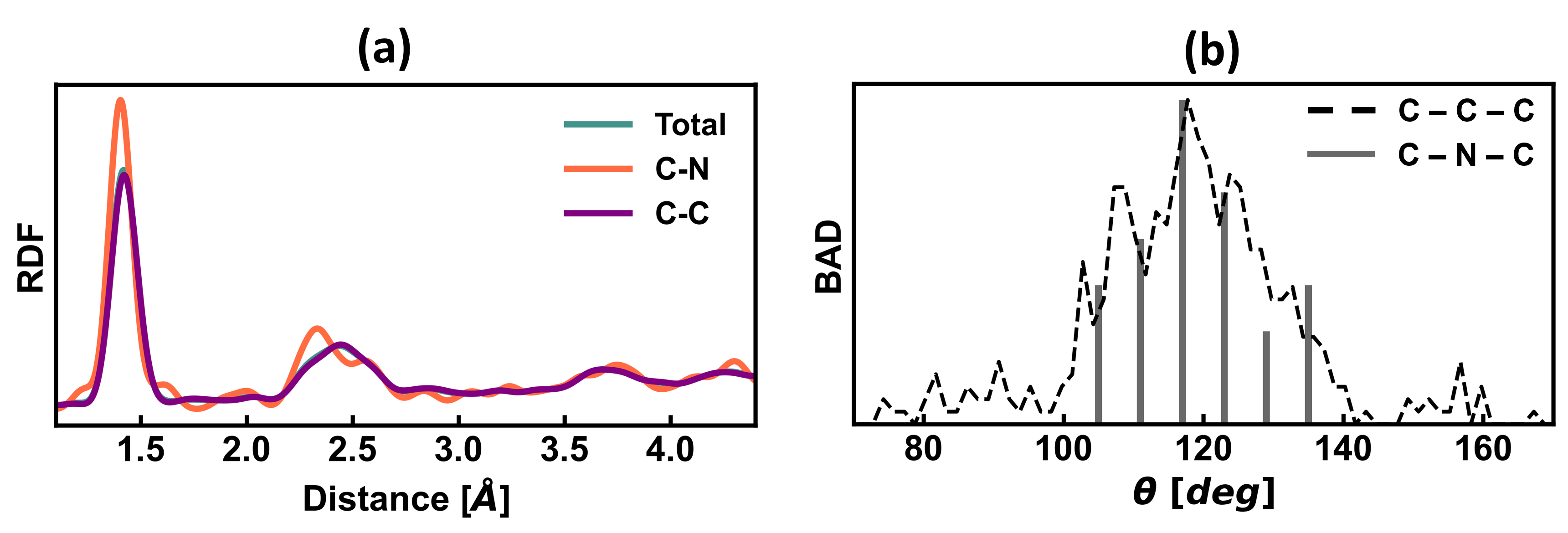}
   
    \caption{(a) illustrates the radial distribution function (RDF) for the nitrogen-containing impure amorphous graphite. Notably, the nearest neighbor distances for carbon-carbon (C--C at 1.419 $\AA$) and carbon-nitrogen (C--N at 1.404 $\AA$) interactions are highly comparable. Bond angle distributions (BAD) for C--C--C and C--N--C are presented in (b). Given the relatively low nitrogen concentration in the model, a histogram was generated for C-N-C angles, depicting angle frequencies. Interestingly, both the central carbon-nitrogen angle and the nitrogen-carbon angle exhibit peak values around 118$^{\circ}$, closely resembling angles in graphite. The BAD plots were normalized to their maximum values. This normalization suggests that nitrogen substitutions within the layered nanostructure maintain the precise sp$^2$ layered arrangement, extending (at least) up to the nearest carbon atom.}
 
	\label{fig:CSfig_RDFBAD_CN}
\end{figure*}

In Figure \ref{fig:CSfig_FullEDoS}a-d, we present the electronic structure of (a) amorphous graphite and impure versions with (b) 5\% and (c) 10\% non-carbon elements. Figure \ref{fig:CSfig_FullEDoS}d illustrates the electronic structure for a layered nanostructure with 5\% nitrogen impurity. A representative model of nitrogen-containing amorphous graphite is shown in Figure \ref{fig:CSfig_FullEDoS}e, and atoms localized at the state with IPR = 0.51 are depicted in Figure \ref{fig:CSfig_FullEDoS}f. Radial distribution function (RDF) and bond angle distribution (BAD) for nitrogen-containing amorphous graphite are presented in Figure \ref{fig:CSfig_RDFBAD_CN}a and b, respectively.

